\documentclass[
superscriptaddress,
 amsmath,amssymb,
 aps,
prstab,
]{revtex4-2}

\usepackage{xcolor}
\usepackage{graphicx}
\usepackage{dcolumn}
\usepackage{bm}
\usepackage[colorlinks=true,citecolor=blue,linkcolor=magenta]{hyperref}
\usepackage{braket}
\def \mc{\mathcal}

\begin{document}

\preprint{APS/123-QED}
\title{Anisotropic Rashba coupling to polar modes in KTaO$_3$}

 \author{Giulia Venditti}
 \affiliation{SPIN-CNR Institute for Superconducting and other Innovative Materials and Devices, Area della Ricerca di Tor Vergata, Via del Fosso del Cavaliere 100, 00133 Rome, Italy}
\author{Maria Eleonora Temperini}%
\affiliation{Department of Physics, Sapienza University of Rome and Istituto Italiano di Tecnologia, Center for Life Nano- \& Neuro-Science, Viale Regina Elena 291, 00161 Rome, Italy}%
\author{Paolo Barone}
\affiliation{SPIN-CNR Institute for Superconducting and other Innovative Materials and Devices, Area della Ricerca di Tor Vergata, Via del Fosso del Cavaliere 100, 00133 Rome, Italy}
\author{Jose Lorenzana}
\affiliation{ISC-CNR Institute for Complex Systems and Department of Physics, Sapienza University of Rome, Piazzale Aldo Moro 2, 00185 Rome, Italy}%
\author{Maria N. Gastiasoro}
\email{maria.ngastiasoro@uniroma1.it}
\affiliation{ISC-CNR Institute for Complex Systems and Department of Physics, Sapienza University of Rome, Piazzale Aldo Moro 2, 00185 Rome, Italy}

\date{\today}
             
\begin{abstract}
Motivated by the discovery of superconductivity in KTaO$_3$-based heterostructures, we study a pairing mechanism based
on spin-orbit assisted coupling between the conduction electrons and the ferroelectric modes present in the material.
We use \emph{ab initio} frozen-phonon computations to show a linear-in-momentum Rashba-like coupling with a strong angular dependence in momentum for the lower $j=3/2$ manifold, deviating from the conventional isotropic Rashba model. This implies the Rashba-like interaction with the polar modes has substantial $L=3$ cubic harmonic corrections, which we quantify for each electronic band. 
The strong anisotropy of the Rashba interaction is captured by a microscopic toy model for the $t_{2g}$ electrons. We find its origin to be the angular dependence in electronic momentum imposed by the kinetic term on the degenerate $j=3/2$ manifold. 
A comparison between the  toy model and \emph{ab initio} results indicate that additional symmetry allowed terms beyond odd-parity spin-conserving inter-orbital hopping processes are needed to describe the Rashba-like polar interaction between the electrons and the soft ferroelectric mode. 
\end{abstract}

\maketitle

\section{Introduction}

Recently a new family of superconductors has been discovered on the interfaces between KTaO$_3$ (KTO) and other oxide insulators, as well as on uncapped KTO surfaces doped with ionic gating~\cite{liu2021two,cheng2021,chen2021electric,ren2022two,liu2022tunable}. The observed superconducting critical temperature  $T_c$ of the two-dimensional electron gas (2DEG) is an order of magnitude higher than the $T_c$  in the closely related SrTiO$_3$ (STO) based heterostructures~\cite{reyren2007}. Therefore insights into the pairing mechanism in these materials are of great interest. 

Remarkably, the $T_c$ shows a strong sensitivity on the crystallographic orientation of KTO~\cite{liu2022tunable}: $T_c=2$K at the (111) interface, $T_c=$1K at the (110) interface and no signal of superconductivity at the (001) interface~\cite{ren2022two,liu2022tunable}. 
It has been proposed~\cite{liu2022tunable} that Cooper pairing mediated by the linear coupling to the soft transverse optical (TO) mode, the so called ferroelectric (FE) mode, can explain the strong directional dependence of $T_c$. The electron-phonon vertex with the soft FE mode was modelled following Ref.~\cite{gastiasoro2022theory}, where a spin-orbit coupling assisted Rashba-like coupling to a polar mode was derived for incipient ferroelectric systems. 
In fact, like the related material STO, the dielectric constant of KTO grows to extremely large values at low $T$~\cite{fujii1976}, staying on the verge of a ferroelectric transition, with a concomitant softening of a TO mode~\cite{Vogt1988}. 
In STO, the Rashba pairing mechanism has been found to have a BCS coupling constant of the right order of magnitude to support bulk superconductivity, and develop a dome of $T_c$ vs carrier density with remarkable experimental agreement~\cite{gastiasoro2022STO}.

Motivated by these promising results and encouraging implications for the Rashba pairing mechanism,  
in this work we present a careful study of the linear coupling between polar modes and the three spin-orbit coupled bands of KTO, going beyond the approximate estimates given in Ref.~\cite{liu2022tunable}. 
Following the approach we developed in Refs.~\cite{gastiasoro2022theory,gastiasoro2022STO} we perform relevant frozen-phonon \emph{ab initio} computations in bulk KTO and extract the odd-in-momentum Rashba-like couplings from the reconstructed electronic band structure. 
We find that besides substantial Rashba couplings to relevant polar modes, a minimal phenomenological form of the interaction goes beyond the isotropic Rashba model for the two lowest electronic bands of the $j=3/2$ manifold, and quantify the anisotropy of the coupling. Moreover, both the strength and anisotropy of the Rashba coupling strongly depend on the particular eigenvector of the polar mode. 

We present a microscopic toy model for $t_{2g}$ electrons coupled to a polar displacement via odd-parity spin conserving inter-orbital hopping processes. The toy model captures many of the frozen-phonon \emph{ab initio} features, including the strong anisotropy (isotropy) of the Rashba coupling in the $j=3/2$ ($j=1/2$) multiplet. 
The angular dependence in the coupling arises from an angular dependent breaking of the degeneracy of the $j=3/2$ states by the hopping terms in the Hamiltonian.

Notably, the strong sensitivity of the Rashba coupling features on the polar eigenvector found by \emph{ab initio} cannot be described by this one-parameter toy model. 
Considering additional symmetry allowed terms besides the odd-parity inter-orbital hopping processes would introduce extra parameters in the model, and may succeed in capturing the polar eigenvector dependent features.

The paper is organized as follows. Section~\ref{sec:electronic} and Section~\ref{sec:Si} introduce the relevant electronic and polar phonon degrees of freedom in bulk KTO, respectively. 
Section~\ref{sec:linear} contains the results on the linear coupling between the electrons and polar modes: the strong anisotropic (isotropic) Rashba-like interaction inferred from frozen-phonon computations in the $j=3/2$ ($j=1/2$) manifold, and a microscopic model that captures many of the essential features of these \emph{ab initio} results.  Section~\ref{sec:conclusions} summarizes the main findings and highlights relevant open questions.

\section{Model parametrization of the DFT electronic structure}
\label{sec:electronic}

\begin{figure}[t]
    \centering
    \includegraphics[width=\linewidth]{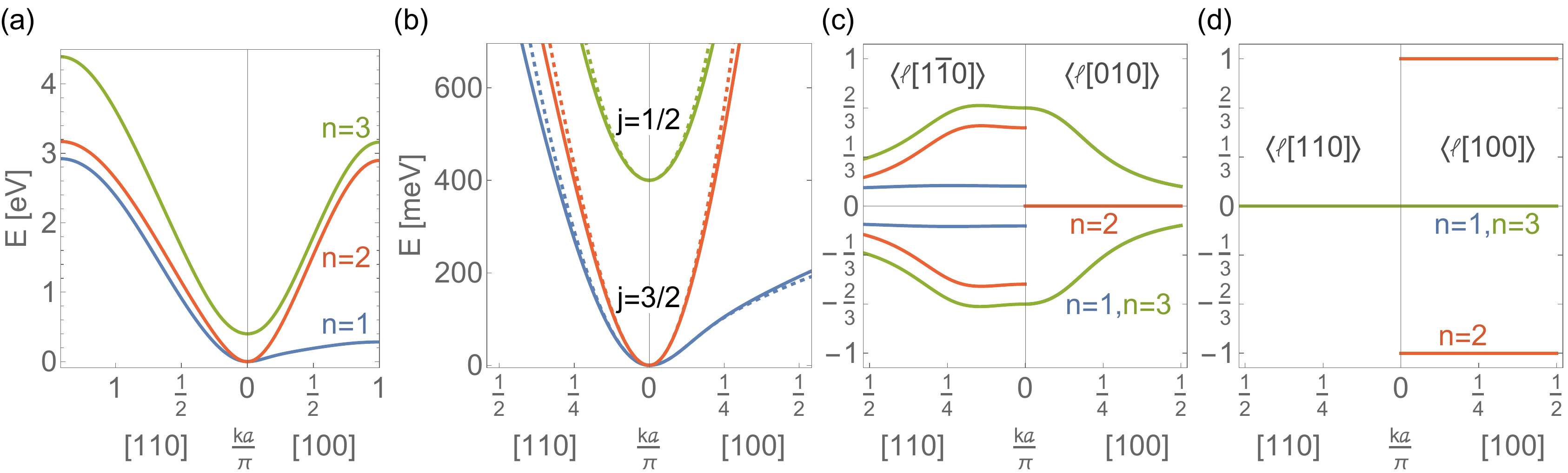}
    \caption{(a) \emph{Ab initio} electronic structure of the three conduction bands in bulk KTO along the $M$ -- $\Gamma$ -- $X$. (b) Zoom into the low-energy dispersion, with the $3\xi=0.4$ eV SOC gap between lower quartet $j=3/2$ [Eqs.\eqref{eq:c3232}-\eqref{eq:c3212}] and upper doublet $j=1/2$ [Eq.~\eqref{eq:c1212}]. Full lines are \emph{ab initio} results and dashed lines the tight-binding model Eq.~\eqref{eq:tb-model}. 
    $a=3.99$ {\AA} is the cubic lattice constant. 
    Orbital angular momentum $\braket{\bm{l}}$ of the electronic bands in the tight-binding model along the same $\bm{k}$ path in the presence of an infinitesimal polarization (see text) for (c) $\bm{l}\perp\bm{\hat{k}}$ and (d) $\bm{l}\parallel\bm{\hat{k}}$. The polar axis $\bm{\hat{n}}_p\parallel[001]$ is perpendicular to both $\bm{l}$ and $\bm{\hat{k}}$ and $\braket{l[001]}=0$. }
    \label{fig:bands}
\end{figure}

The electronic band structure of bulk KTO computed by \emph{ab initio} is shown in Fig.~\ref{fig:bands}(a), within the energy window of the three doubly degenerate conduction bands, which we denote $n=1$, $n=2$ and $n=3$ from lowest to highest (see also \ref{app:dft} for computational details). 

The low-energy part of the dispersion from \emph{ab initio}, shown in Fig.~\ref{fig:bands}(b) by solid lines, can be effectively described by a tight-binding model of the Ta $5d$ electrons supplemented with an atomic spin-orbit interaction. That is, a model including the $yz$, $zx$ and $xy$ orbitals within the $t_{2g}$ manifold, referred to as $\mu=x,y,$ and $z$ respectively in this work. 
This non-interacting Hamiltonian is given by
\begin{align}
\label{eq:tb-model}
    \mc{H}=\mc{H}_0+\mc{H}_\mathrm{SOC}&= \sum_{n  \bm{k} }  \psi_n^\dagger(\bm{k}) \mc E_{n}(\bm{k}) \sigma_0 \psi_{n}(\bm{k})   \\
    \label{eq:H0}
    \mc{H}_0&=\sum_{\bm{k} s \mu\nu}t_{\mu\nu}(k)c^\dagger_{\mu s}(\bm{k}) c_{\nu s}(\bm{k})\\
    \mc{H}_\mathrm{SOC}&=\xi\sum_{\bm{k}\mu s\nu s' l}i \epsilon_{\mu\nu l}\sigma_{\bm{k}l,ss'}c^\dagger_{\mu,s}(\bm{k})c_{\nu,s'}(\bm{k}).
    \label{eq:Hsoc}
\end{align}
It includes a hopping term $\mc{H}_0$ between orbitals $\mu$ and $\nu$ with spin-$\frac{1}{2}$ ($s=\pm$) up to next-nearest neighbors,
\begin{align}
     t_{\mu\mu}(\bm{k})=&-2t_1\left(\cos k_\alpha+ \cos k_\beta \right)-2 t_2 \cos k_\mu -4 t_3 \cos k_\alpha \cos k_\beta+(4t_1+2t_2+4t_3)\label{eq:tmumu}\\
   t_{\mu\nu}(\bm{k})=&-4 t_4 \sin k_\mu \sin k_\nu,\label{eq:tmunu}
 \end{align}
with hopping parameters $t_1=526$ meV, $t_2=33$ meV, $t_3=214$ meV and $t_4=30$ meV. In Eq.~\eqref{eq:tmumu} $\alpha\neq\beta\neq\mu$, while in Eq.~\eqref{eq:tmunu},  $\mu \neq \nu$. 

The $\mc{H}_\mathrm{SOC}$ term in Eq.~\eqref{eq:tb-model} describes the atomic SOC of the $t_{2g}$ manifold with the conventional effective orbital moment $l=1$~\cite{khomskii2020orbital}, the Levi-Civita symbol $\epsilon_{\mu\nu l}$ and Pauli matrix $\bm{\sigma}$. At the zone-center, where the hopping term  $\mc{H}_0$ is zero, the mixing of the orbital and spin degrees of freedom of $\mc{H}_\mathrm{SOC}$ breaks the sixfold degeneracy of the $t_{2g}$ manifold. This results in the following set of eigenstates at the zone-center, 
\begin{align}
\label{eq:c3232}
    c^\dagger_{\frac{3}{2},\pm\frac{3}{2}}&=\mp\frac{1}{\sqrt{2}}\left(c^\dagger_{x,\pm}\pm i c^\dagger_{y,\pm}\right) 
    \\
    \label{eq:c3212}
    c^\dagger_{\frac{3}{2},\pm\frac{1}{2}}&=\frac{1}{\sqrt{6}}\left(\mp c^\dagger_{x,\mp}- i c^\dagger_{y,\mp}+ 2 c^\dagger_{z,\pm}\right) 
    \\
    c^\dagger_{\frac{1}{2},\pm\frac{1}{2}}&=\frac{1}{\sqrt{3}}\left(-c^\dagger_{x,\mp}\mp i c^\dagger_{y,\mp}\mp  c^\dagger_{z,\pm}\right) 
    \label{eq:c1212}
\end{align}
where the $j=1/2$ doublet is pushed $3\xi=400$ meV up in energy with respect to the $j=3/2$ quartet. This gap can be seen in Fig.~\ref{fig:bands}(b), and, as previously pointed out~\cite{bruno2019band}, it is an order of magnitude larger than the 28 meV SOC gap of cubic STO, a close analog material with conduction bands formed of $3d$ electrons of the Ti atoms instead. 

For a general momentum $\bm{k}$, where both $\mc{H}_0$ and $\mc{H}_\mathrm{SOC}$ are finite, the electronic spinor $\psi^\dagger_n(\bm{k})=(c^{\dagger}_{n+}(\bm{k}),c^{\dagger}_{n-}(\bm{k}))$ of band $n=1,2,3$ in Eq.~\eqref{eq:tb-model} develops a doubly degenerate dispersion $E_n(\bm{k})$ in pseudospin (described by the $2\times2$ identity matrix $\sigma_0$). The resulting band structure of this tight-binding+SOC model with the parameters $t_1, t_2, t_3, t_4$ and $\xi$ quoted above is shown in Fig.~\ref{fig:bands}(b) (dashed lines), showing an excellent agreement with the \emph{ab initio} computation (full lines).

The eigenstates of  $\mathcal{H}_0$  [Eq.~\eqref{eq:H0}] are degenerate at $\Gamma$. This degeneracy is removed by the 
SOC which favours eigenstates with finite orbital angular momentum  Eqs.~\eqref{eq:c3232}-\eqref{eq:c1212}. Their residual degeneracy, however, will be affected at finite $k$ by the effective mass mismatch of the different bands described by  $\mc{H}_0$.
For instance, the additional degeneracy of the $j=3/2$ multiplet at the zone-center is broken by the cubic hopping term  $\mathcal{H}_0$  in a $\bm{\hat{k}}$-dependent manner, which can be shown applying degenerate perturbation theory. We anticipate that how the degeneracy of the manifold is broken by  $\mathcal{H}_0$ will become important when discussing the origin of the anisotropic (isotropic) interaction of the $j=3/2$ ($j=1/2$) manifold with polar modes in Section~\ref{subsec:origin}.   
The states in 
Eqs.~\eqref{eq:c3232}-\eqref{eq:c1212}, have a finite orbital angular momentum along $z$. In cubic symmetry at $\Gamma$ this is completely arbitrary as one can choose any direction as the quantization axis. For finite 
momentum, the effect of  $\mathcal{H}_0$ partially removes this freedom.  
Still, 
even when the degeneracy is reduced to two fold (Kramers degeneracy) there is some degree of arbitrariness on the matrix elements of the angular momentum, due to the arbitrary choice of pseudospin basis. One can remove this arbitrariness by adding an infinitesimal polarization [Fig.~\ref{fig:bands}(c)-(d)]. These results will become useful later to understand the dependence of the Rashba-like coupling on the electronic momentum, both in magnitude and direction  (Sec.~\ref{subsec:origin}).

An analogous tight-binding model for cubic STO~\cite{gastiasoro2022theory} found very similar hopping parameters $t_i$ [Eqs.~\eqref{eq:tmumu}-\eqref{eq:tmunu}] but, in line with the much smaller SOC gap in that material, a proportionally smaller SOC parameter: $\xi^\mathrm{KTO}=14.5\xi^\mathrm{STO}$. This will have important effects for the momentum dependence of the linear polar coupling and the validity of the conventional linear-in-$k$ Rashba regime, as we will explain in Section~\ref{subsec:origin}.

\section{Polar soft mode in KTO}
\label{sec:Si}

The atomic displacements of the zone-center polar soft mode of KTO can be given in terms of a complete set of symmetry coordinates for the $T_{1u}$ irrep of $O_h$. 
A suitable set of coordinates apt to describe the eigenvectors of polar normal modes follows from cubic perovskites~\cite{1967Axe}.
For a generic polar axis $\bm{\hat{n}}_p$
three $\bar S_i$ symmetrized modes are 
defined: ``Slater", with a vibration of the Ta against the oxygen cage ($i=1$), the vibration of the potassium out of phase with a rigid TaO$_3$ cage ($i=2$) and the deformation of the oxygen cage ($i=3$). 
Then an arbitrary polar distortion $\bm{ \bar U}$ 
along that polar axis can be decomposed into the three symmetrized polar modes $\bm{ \bar U}=\bm{\hat{n}}_p\sum_i u_i \bar S_i$. Unlike lower symmetry STO\cite{gastiasoro2022theory},  all displacements are in the same direction of the polar axis.  Here, $u_i$ describes the amplitude of symmetrized mode $i$
and $ \bar S_i$ defines the symmetrized mode in terms of normalized atomic displacements $(s^\mathrm{K},s^\mathrm{Ta},s^{\mathrm{O}_x},s^{\mathrm{O}_y},s^{\mathrm{O}_z})$,  
\begin{align}
\label{eq:S1}
   \bar S_1&=\frac{1}{1+\kappa_1}(0,-\kappa_1,1,1,1)  \\
  \label{eq:S2}
    \bar S_2&=\frac{1}{1+\kappa_2}(-\kappa_2,1,1,1,1) \\
  \label{eq:S3}
   \bar S_3&=\frac{2}{3}(0,0,-\frac{1}{2},-\frac{1}{2},1)
\end{align}
where the coefficients $\kappa_1=\frac{3m^\mathrm{O}}{m^\mathrm{Ta}}$ and $\kappa_2=\frac{3m^\mathrm{O}+m^\mathrm{Ta}}{m^\mathrm{K}}$ ensures that the center of mass is not displaced for any of the $\bar{S_i}$ modes. 
That is, $\kappa_1$ and $\kappa_2$ assures the condition $\sum_l m^l \bm{r}^l = 0$ to be fulfilled, the sum running over all the $l$ atoms having atomic mass $m^l$ and displacement $r^l=u_is^l$ in the unit cell for each of the $\bar S_i$ modes.
The normalization of $\bar S_i$ modes in Eqs.~\eqref{eq:S1}-\eqref{eq:S3} are set so that $u_i$ is the relative displacement of the two bodies in the mode. 
Hence, when deriving the electron-phonon Hamiltonian in Section~\ref{subsec:el-ph}, the two-body problem will be reduced to a one body problem with a reduced mass.
For instance, the relative atomic displacement $u_1$ between Ta and the O cage in the 
 $\bar S_1$ mode is set to be $r^{\mathrm{O}}-r^\mathrm{Ti}=u_1(s_1^\mathrm{O}-s_1^\mathrm{Ti})=u_1$. 
Similarly, $u_2$ ($u_3$) is the relative displacement between K (O$_z$) and the Ta-O cage (O$_x$, O$_y$) in mode $\bar S_2$ ($\bar S_3)$. 
The bar symbol indicates a vector spanned by the atoms of the unit cell (as in $\bar S_i$), whereas the bold notation (as in $\bm{\hat{n}}_p$) refers to the  Cartesian coordinates of the atomic displacements.

In this work, we will explore distortions of the cubic structure with polar axis $\bm{\hat{u}}$ along the $[001]$ direction for the $\bar S_i$ modes. 
While the soft TO mode in bulk KTO is in general a linear combination of all three $\bar S_i$ modes~\eqref{eq:S1}-\eqref{eq:S3}, according to neutron scattering and hyper-Raman experiments the $\bar S_1$ mode has the dominant contribution~\cite{Harada1970, Vogt1988}.
Moreover, since we are interested in the coupling of the electronic bands to the polar modes to linear order, we consider only displacements small enough to stay in the linear regime near the $\Gamma$ point (for details see~\ref{app:u*}).

\section{Linear polar coupling}
\label{sec:linear}

\subsection{Anisotropic Rashba coupling from DFT}
\label{subsec:anisoRashba}

The linear coupling between a polar mode $\bm{u}_i(\bm{q})=u_i(\bm{q})\bm{\hat{n}}_p(\bm{q})$ described in Section~\ref{sec:Si} and the SOC electrons in Eq.~\eqref{eq:tb-model} takes the following form:
\begin{equation}
 \mc{H}_u= \sum_{n \bm{k} \bm{q},\bar S_i }  \psi_n^\dagger(\bm{k}+\frac{\bm{q}}{2}) \bm{\Lambda}^{\bar S_i}_{n}(\bm{k},\bm{q}) \psi_{n}(\bm{k}-\frac{\bm{q}}{2})  
 \label{eq:Husoc}
\end{equation}
with the coupling $ 2\times 2$ matrix $\bm{\Lambda}^{\bar S_i}_{n}(\bm{k},\bm{q}) $ in pseudospin space for a mode $\bar S_i$. Note that we are only considering intra-band coupling for each electronic band $n$, and we are ignoring inter-band processes in this work for simplicity. 
We start with the polar coupling matrix to lowest order in electronic momentum $\bm{k}$ allowed by symmetry in a cubic point group~\cite{kozii2015odd,gastiasoro2020,sumita2020}, 
\begin{equation}
\label{eq:Lambdaiso}
    \bm{\Lambda}^{\bar S_i}_{n}(\bm{k},\bm{q}) = \tau^{\bar S_i}_n k a    \bm{\hat{k}} \times \bm{\sigma}\cdot \bm{u_i(\bm{q})},
\end{equation}
linear in both $\bm{k}$ and $\bm{u}_i(\bm{q})$, with the bare coupling constant $\tau^{\bar S_i}_{n}$ for each electronic $n$ band with a polar mode $\bar{S}_i$ [Eqs.~\eqref{eq:S1}-\eqref{eq:S3}]. We discuss briefly the symmetries involved which will be useful below to generalize to higher order. 
Because we consider $T_{1u}$ polar phonons $\bm{u_i(\bm{q})}$, the electronic part of Eq.~\eqref{eq:Lambdaiso} involving momentum and pseudospin has to transform also as $T_{1u}$, and be time reversal symmetric. The lowest order term, $L=1$, in the $L$th multipole expansion is $\bm{\hat{k}} \times \bm{\sigma}$ ($\bm{\hat{k}}$ and $\bm{\sigma}$ transform as $T_{1u}$ and $T_{1g}$, respectively)~\cite{Dresselhaus2008}. 

Equation~\eqref{eq:Lambdaiso} couples the polar distortion $\bm{u}_i(\bm{q})$ to the pseudospin $ \bm{\sigma}$ of the electronic band $n$ in a $k$-linear form reminiscent of the Rashba interaction appearing in polar structures. The main difference is that here the structure is non-polar and the polarization,  $\bm{u_i(\bm{q})}$, is a dynamical field.

We can characterize the coupling $ \tau^{\bar S_i}_n $ by the effect that a $\bm{q}=0$ out-of-equilibrium $\bm{u_i}$ produces on the electronic structure. In such a ``frozen phonon" scheme a characteristic linear-in-$k$ electronic band splitting is induced in the pseudospin sector, $E_{n,+}(\bm{k})-E_{n,-}(\bm{k})=\delta E_n(\bm{k})$ where 
$\pm$ refer to the pseudospin quantum number along the quantization axis defined by the $\bm{\hat{n}}_p \times \bm{\hat{k}}$ direction.
The splitting reads, 
\begin{equation}
\label{eq:dEiso}
    \delta E_n(\bm{k},\bm{u}_i(\bm{0}))=2\tau^{\bar S_i}_n k a u_i(0) |\bm{\hat{n}}_p \times \bm{\hat{k}}|.  
\end{equation}
This band split is maximum (zero) when the electronic momentum $\bm{k}$ is perpendicular (parallel) to the polar axis $\hat{\bm{n}}_p$ of the mode. Eq.~\eqref{eq:dEiso} is also isotropic along the azimuthal angle of the polar axis $\hat{\bm{n}}_p$, i.e. it has axial symmetry around $\hat{\bm{n}}_p$. 

Following Refs.~\cite{gastiasoro2022theory,gastiasoro2022STO} we will now use the electronic band split computed from frozen-phonon \emph{ab initio} to extract the Rashba-like couplings $\tau_{n}$ of the three electronic bands of KTO [Fig.~\ref{fig:bands}(b)]. 
For a mode with the polar axis along $\hat{\bm{n}}_p\parallel [001]$ the electronic split in Eq.~\eqref{eq:dEiso} allowed by symmetry becomes
\begin{equation}
\label{eq:dEisonz}
    \frac{\delta E_n(\bm{k},u_i \hat{\bm{n}}_p)}{2u_i}=ka\tau^{\bar S_i}_{n}|\sin{\theta}|
\end{equation}
where we have introduced the polar angle $\theta$ of the momentum direction $\bm{k}$ measured from the polar axis $\hat{\bm{n}}_p\parallel [001]$. The axial symmetry of the split in the perpendicular $\bm{k}$ direction ($\theta=\frac{\pi}{2}$) is shown in Fig.~\ref{fig:K3}(d), i.e. the conventional circular Rashba field.

\begin{figure}[t]
    \centering 
    \includegraphics[width=\linewidth]{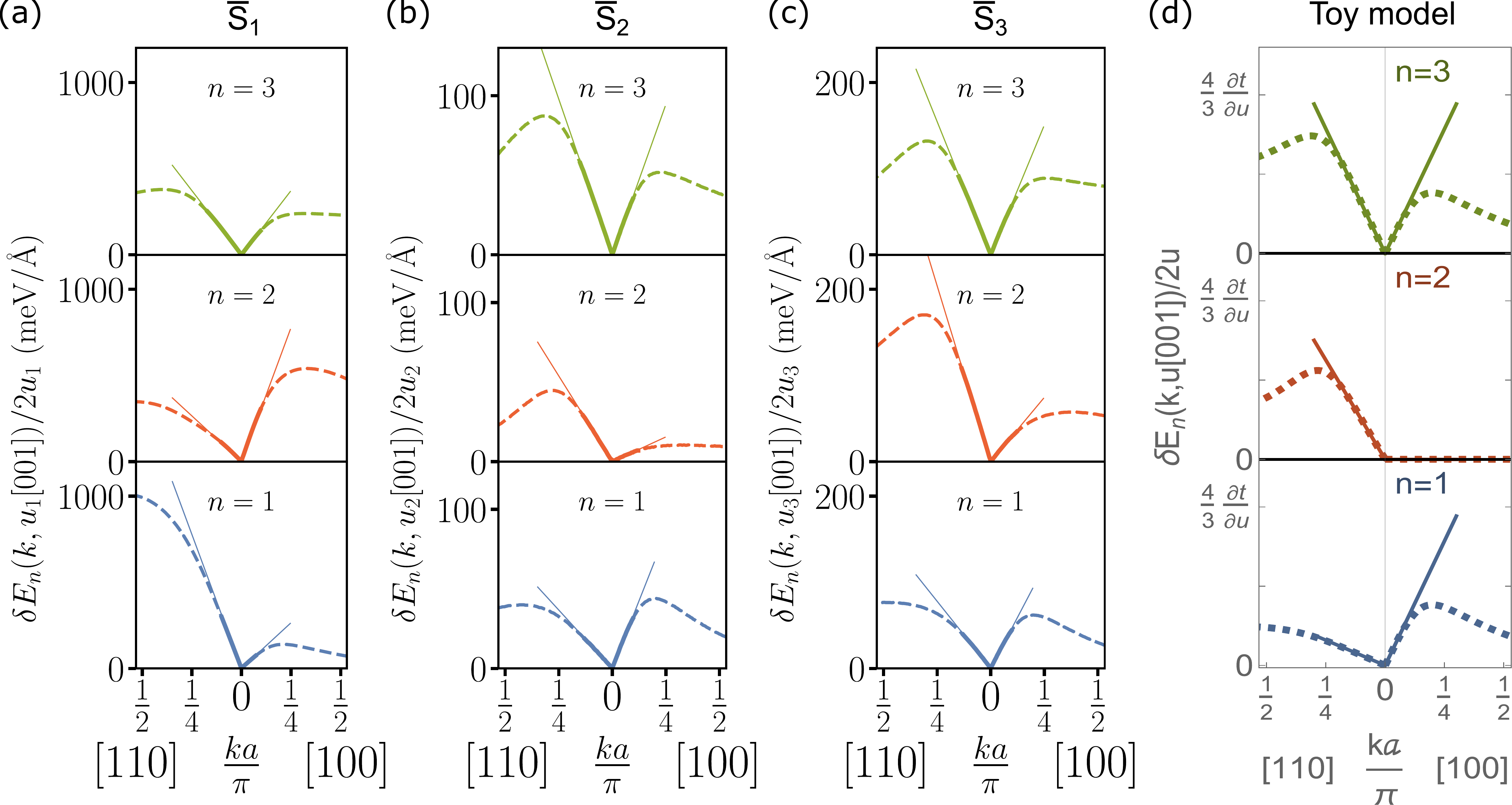}
    \caption{Electronic band split $|\delta E_{n}(\bm{k},\bm{u}_i(q=0)\bm{\hat{n}}_p)|$ in the presence of a frozen polar mode with $\bm{\hat{n}}_p \parallel[001]$, normalized by twice its amplitude $2u_i$ for
    modes (a) $\bar S_1$, (b) $\bar S_2$ and (c) $\bar S_3$ computed by  \emph{ab initio} (dashed lines) and (d) computed by  microscopic toy model Eq.~\eqref{eq:Htotmodel} (dashed lines) for a generic polar distortion belonging to the $T_{1u}$ irrep. 
    Full lines show the $k$-linear Rashba split in (a)-(c) Eqs.~\eqref{eq:dE100}-\eqref{eq:dE110} and (d) Eqs.~\eqref{eq:dE1}-\eqref{eq:dE3} up to $\frac{ka}{\pi}=0.3$.
    Each vertical panel corresponds to electronic band $n=1$ (blue), $n=2$ (red) and $n=3$ (green).}
    \label{fig:linearfit}
\end{figure}

The reconstructed electronic band structure of bulk KTO in the presence of a frozen phonon $\bm{u}_i(\bm{q}=\bm{0})$ with polar axis along $\hat{\bm{n}}_p\parallel[001]$ was then computed by DFT for the three relevant $\bar{S}_i$ modes presented in Section~\ref{sec:Si} [Eqs.~\eqref{eq:S1}-\eqref{eq:S3}]. The resulting electronic band splitting is shown in Figs.~\ref{fig:linearfit}(a)-(c) for each band along two electronic momentum directions perpendicular to the polar axis $\hat{\bm{n}}_p$, $\hat{\bm{k}}\parallel [110]$ and $\hat{\bm{k}}\parallel [100]$. 
As expected, the band-split grows in a linear-in-$k$ conventional Rashba fashion for all bands and modes.

In principle, according to the isotropic form of Eq.~\eqref{eq:dEisonz}, fitting the linear-in-$k$ regime along any given momentum in the $k_x k_y$-plane ($\theta=\pi/2$) should give the same coupling $\tau_{n}$ [see also Fig.~\ref{fig:K3}(d)]. In other words, the linear-in-$k$ slope should not depend on momentum direction $\bm{\hat{k}}$. 
However, by looking at Figs.~\ref{fig:linearfit}(a)-(c), only the split of the highest band $n=3$ of the $j=1/2$ doublet (green data) is well described by this isotropic form, and hence by the minimal polar interaction given by Eq.~\eqref{eq:Lambdaiso}. The lowest two bands $n=1$ (blue) and $n=2$ (red) of the $j=3/2$ multiplet on the contrary, show a large band split anisotropy manifested by the very different slopes displayed along $\hat{\bm{k}}\parallel [110]$ and $\hat{\bm{k}}\parallel [100]$ (also listed in Table~\ref{tab:taudft}), and hence not captured by the isotropic expression in Eq.~\eqref{eq:Lambdaiso}. In addition, the split anisotropy strongly depends on the polar mode $\bar S_i$ for both $n=1$ and $n=2$ electronic bands. 

The frozen-phonon DFT results suggest that one needs to go beyond the conventional isotropic Rashba form of the polar interaction Eq.~\eqref{eq:Lambdaiso} for the lowest two bands $n=1$ and $n=2$ of KTO. We thus allow for angular corrections of the electronic momentum $\bm{\hat{k}}$ in the interaction in the presence of a cubic crystal field, 
\begin{align}
\label{eq:Lambda}
    &\bm{\Lambda}^{\bar S_i}_{n}(\bm{k},\bm{q})=  k a u_i (\bm{q}) \left(\tau^{\bar S_i}_{n,1}\bm{\hat{k}}+\tau^{\bar S_i}_{n,3}\bm{\mathcal{K}}_3(\bm{\hat{k}})\right)\times\bm{\sigma}\cdot \hat{\bm{n}}_p(\bm{q})\\
&\bm{\mathcal{K}}_3(\bm{\hat{k}})=\left(\hat{k}_x\left(2\hat{k}_x^2-3\hat{k}_y^2-3\hat{k}_z^2\right),\hat{k}_y \left(2\hat{k}_y^2-3\hat{k}_z^2-3\hat{k}_x^2\right),\hat{k}_z \left(2\hat{k}_z^2-3\hat{k}_x^2-3\hat{k}_y^2\right)\right)
\end{align}
by including the contribution of the next allowed order cubic harmonic $\bm{\mathcal{K}}_3(\bm{\hat{k}})$ (with $L=3$)
of the $T_{1u}$ irrep~\cite{muggli1972cubic}. 
A non-zero $\tau^{\bar S_i}_{n,3}$ takes the angular dependence of the polar interaction in Eq.~\eqref{eq:Lambda} beyond the standard $L=1$ order assumed in Eq.~\eqref{eq:Lambdaiso}. Note that we are still keeping the linear order for the modulus of the momentum $k$ and polar displacement $\bm{u}_i$ in the interaction (in agreement with the frozen phonon results in Fig.~\ref{fig:linearfit}).

The expression for the new electronic band split $\delta E_n(\bm{k},\bm{u}_i)$ in the presence of a polar displacement incorporates now the cubic angular corrections of the interaction.
As we did above, we set the polar axis along $\hat{\bm{n}}_p\parallel[001]$ and obtain the following expression for the band split in the $\bm{k}$-plane perpendicular to $\bm{\hat{n}}_p$,
\begin{equation}
\label{eq:dEphi}
    \frac{\delta E_n (k(\cos{\varphi},\sin{\varphi},0),u_i[001])}{2k a u_i}=\tau^{\bar S_i}_{n,1}\sqrt{1+\frac{\tau^{\bar S_i}_{n,3}}{\tau^{\bar S_i}_{n,1}}\left(a_0+b_0\cos{4\varphi}\right)}
\end{equation}
with coefficients $a_0=\frac{3}{2}+\frac{17}{8}\frac{\tau^{\bar S_i}_{n,3}}{\tau^{\bar S_i}_{n,1}}$ and $b_0=\frac{5}{8}\left(4+3\frac{\tau^{\bar S_i}_{n,3}}{\tau^{\bar S_i}_{n,1}}\right)$.
The band split in Eq.~\eqref{eq:dEphi} has now two extrema along the $\bm{\hat{k}}\parallel [100]$ and $\bm{\hat{k}}\parallel [110]$ directions: 
\begin{align}
\label{eq:dE100}
    \frac{\delta E_n (\varphi=0)}{2k a u_i}&=\tau^{\bar S_i}_{n,1}+2\tau^{\bar S_i}_{n,3}\\
    \frac{\delta E_n (\varphi=\pi/4)}{2k a u_i}&=\tau^{\bar S_i}_{n,1}-\frac{1}{2}\tau^{\bar S_i}_{n,3}
    \label{eq:dE110}
\end{align}
That is, the band split is now anisotropic in the $\hat{k}_x\hat{k}_y $-plane for a finite $\tau^{\bar S_i}_{n,3}$, the axial symmetry broken. 
Indeed, Eq.~\eqref{eq:dEphi} manifests the cubic angular correction expected from an object belonging to the $A_{1g}$ irrep, i.e. $\mathcal{F}_{A_{1g}}(\theta,\varphi)\approx c_0+c_4K_4(\theta,\varphi)$ with the 
$L=4$ cubic harmonic taking the form $K_4(\theta=\pi/2,\varphi)= 1+\frac{5}{3}\cos{4\varphi}$ in the equatorial plane.

\begin{figure}[t]
    \centering
    \includegraphics[width=0.75\linewidth]{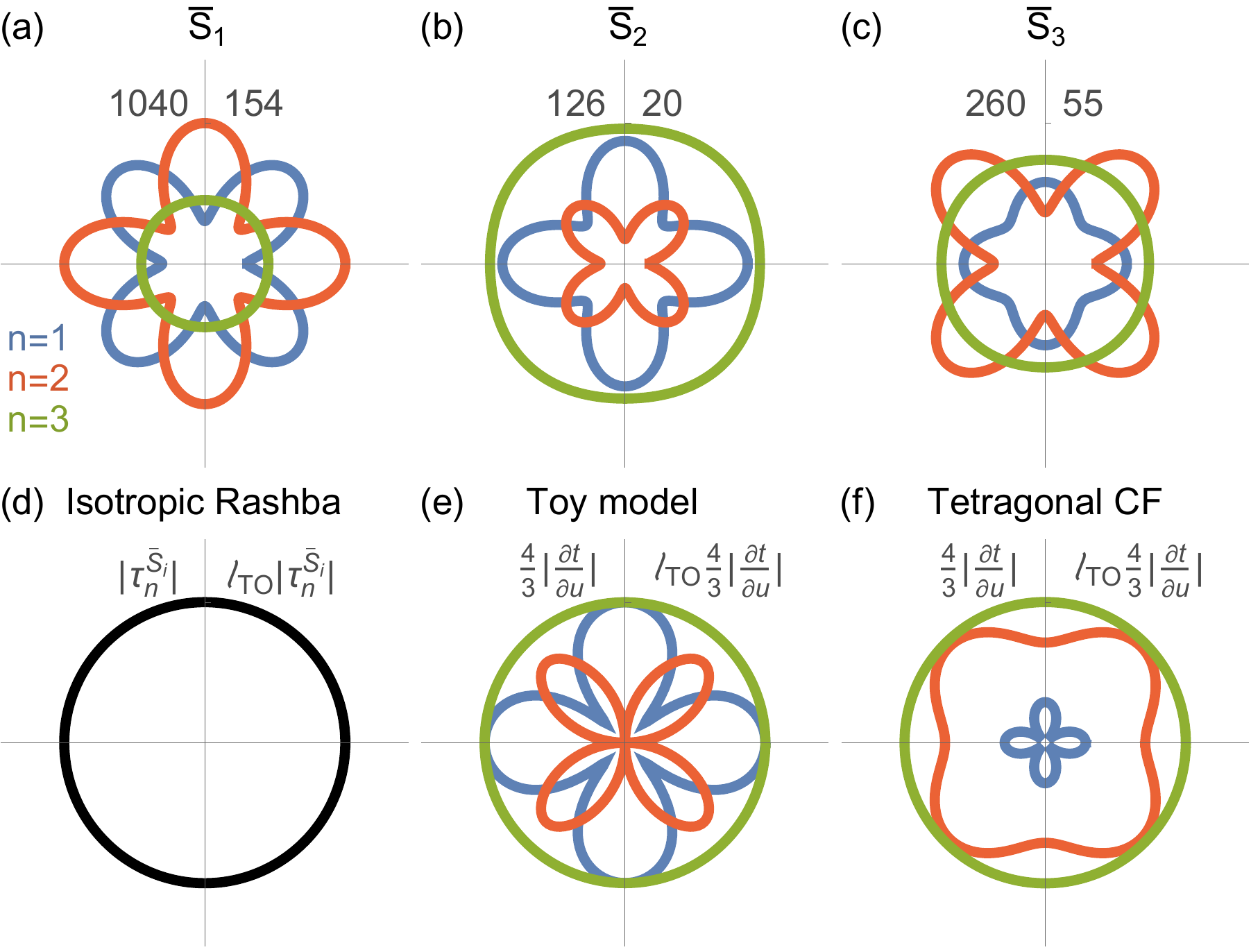}
    \caption{Polarplot of the electronic band split $\frac{|\delta E_n(\bm{k},\bm{u}_i)|}{2kau_i}$ induced by a polar mode with amplitude $u_i$ and polar axis $\bm{\hat{n}}_p\parallel [001]$ in the perpendicular momentum plane $\bm{\hat{k}}=(\cos{\varphi},\sin{\varphi},0)$. The $y$-axis ticks indicate the  value of the largest split (left, meV/\AA) and the corresponding electron-phonon matrix element ${\mathfrak{g}^\mathrm{TO}_n(\bm{k})}/{ka}$ (right, meV).  
    (a)-(c) Anisotropic Rashba including $L=3$ cubic corrections [Eq.~\eqref{eq:dEphi}], with $\tau^{\bar S_i}_{1,n}$, $\tau^{\bar S_i}_{3,n}$ obtained from linear fits to frozen-phonon computations in KTO [Fig.~\ref{fig:linearfit}(a)] for a mode (a) $\bar{S}_1$ (Slater), (b)  $\bar{S}_2$ and (c)  $\bar{S}_3$. The split of each electronic band is represented by the same color scheme: $n=1$ (blue) $n=2$ (red) and $n=3$ (green). 
    (d) Conventional isotropic Rashba [Eq.~\eqref{eq:dEisonz}] with a coupling $\tau_n^{\bar S_i}$.
    (e) Microscopic toy model Eq.~\eqref{eq:Htotmodel} with induced inter-orbital hopping amplitude $\frac{\partial t}{\partial u}$ at $\frac{ka}{\pi}=0.01$.
    (f) Same as (e) but with additional term in Eq.~\eqref{eq:Htet} which breaks the degeneracy of $j=3/2$ multiplet.
    }
    \label{fig:K3}
\end{figure}

Using the expressions Eqs.~\eqref{eq:dE100}-\eqref{eq:dE110} to fit the \emph{ab initio} frozen-phonon results of Fig.~\ref{fig:linearfit}(a)-(c), we obtain the Rashba couplings $\tau^{\bar S_i}_{n,1}$ and $\tau^{\bar S_i}_{n,3}$. These in turn describe the minimal Rashba-like polar interaction Eq.~\eqref{eq:Lambda} for the three electronic bands with each polar mode $\bar S_i$. We list the estimated couplings in Table~\ref{tab:coeffs}, and show the resulting electronic band split for each of the three $\bar{S}_i$ modes in Figs.~\ref{fig:K3}(a)-(c). Comparing these results to the isotropic form in Fig.~\ref{fig:K3}(d), we conclude that only the $n=3$ band (green line) can be approximated by the simple isotropic interaction in Eq.~\eqref{eq:Lambdaiso}. The band split of the lowest two bands $n=1$ and $n=2$ shows a pronounced anisotropy (blue and red curves); hence a minimal Rashba-like interaction which includes the $L=3$ harmonic correction in Eq.~\eqref{eq:Lambda} is more appropriate for the bands stemming from the $j=3/2$ multiplet.

\begin{table}
 \caption{(a) Rashba couplings of the cubic expansion $\tau^{\bar S_i}_{n,1} $ and $\tau^{\bar S_i}_{n,3}$ appearing in the polar interaction Eq.~\eqref{eq:Lambda}. Obtained from linear-in-$k$ fits to $\bar{S}_i$ frozen-phonon \emph{ab initio} computations shown in Fig.~\ref{fig:linearfit}(a)-(c), in meV/{\AA}. (b) Corresponding electron-phonon matrix-elements (in meV) in Eq.~\eqref{eq:gnTO}, assuming a frequency $\omega_\mathrm{TO}=2.5$ meV of the soft FE mode~\cite{vogt1995}.  }
     \begin{tabular}{c|ccc|ccc|ccc}
        & & $\bar S_1$ & & & $\bar S_2$ & & & $\bar S_3$ & \\
          &  $n=1$ & $n=2$ & $n=3$ & $n=1$ & $n=2$ & $n=3$ &$n=1$ & $n=2$ & $n=3$\\
         \hline
         $|\tau^{\bar S_i}_{n,1}|$ & 815 & 496 & 478 & 61 & 59 & 125& 109 & 227 & 196\\
         $\frac{\tau^{\bar S_i}_{n,3}}{\tau^{\bar S_i}_{n,1}}$ & -0.31 & 0.55 & -0.008 & 0.40 & -0.31 & -0.016 & 0.19 & -0.29 & -0.01\\
         \hline
         $\frac{|g^{\bar S_i}_{n,1}|}{ka}$ & 121 & 74 & 71 & 10 & 9 & 20 & 31 & 64 & 55\\
         $\frac{g^{\bar S_i}_{n,3}}{g^{\bar S_i}_{n,1}}$ & -0.31 & 0.55 & -0.008 & 0.40 & -0.31 & -0.016 & 0.19 & -0.29 & -0.01
    \end{tabular}
    \label{tab:coeffs}
\end{table}

\subsection{Origin of anisotropic Rashba from a microscopic toy model}
\label{subsec:origin}

We now illustrate how the anisotropy of the Rashba split in the lower $j=3/2$ multiplet naturally emerges when considering microscopic processes induced by a polar displacement~\cite{Petersen2000,Khalsa2013,Zhong2013,Djani2019,gastiasoro2022theory,kumar2022spin,gastiasoro2022STO} and applying degenerate perturbation theory for the hopping term Eq.~\eqref{eq:H0}.

In order to understand the origin of the band splittings in frozen phonon computations we consider a uniform polar distortion ($\bm{q}=0$) along $\bm{\hat{n}}_p\parallel[001]$ with amplitude $u$. This leads to odd-in-$k$ spin-conserving inter-orbital hopping channels~\cite{gastiasoro2022STO} forbidden in the non-interacting Hamiltonian of the $t_{2g}$ manifold Eq.~\eqref{eq:tb-model}, but allowed in the presence of a polar displacement. 
These induced inter-orbital processes have been found very relevant in STO-based heterostructures~\cite{Khalsa2013,Zhong2013} and tetragonal STO~\cite{gastiasoro2022STO}. In order to understand their role here we consider a toy model, restricting to spin conserving processes. The resulting microscopic polar Hamiltonian in the $t_{2g}$ manifold to linear order in the polar displacement reads,
\begin{align}
\label{eq:Huwt}
&\mathcal{H}^\mathrm{mic}_u=\sum_{\bm{k}}\sum_{\mu=x,y}\psi^\dagger_{\mu}(\bm{k})t_{\mu z}(\bm{k},0)\sigma_0 \psi_{z}(\bm{k})+\mathrm{h.c.}\\
&t_{\mu z}(\bm{k},0)=2i\frac{\partial t}{\partial u }u \sin(k_\mu a)
\end{align}
in terms of the spinor of the $t_{2g}$ orbitals $\psi^\dagger_\mu=(c^{\dagger}_{\mu+},c^{\dagger}_{\mu-})$, inter-orbitally connected by an induced hopping amplitude $\frac{\partial t}{\partial u }$ along bonds perpendicular to the polar axis $[001]$. 
Together with the  non-interacting Hamiltonian for the $t_{2g}$ sector Eq.~\eqref{eq:tb-model} we obtain the following toy model which now includes the interaction of the $t_{2g}$ electrons with the polar displacement through Eq.~\eqref{eq:Huwt},
\begin{equation}
\label{eq:Htotmodel}
\mathcal{H}=\mathcal{H}_0+\mathcal{H}_\mathrm{SOC}+\mathcal{H}_u^\mathrm{mic}.
\end{equation}
The resulting electronic band split for each band for a finite hopping amplitude $\frac{\partial t}{\partial u }$ in the $\bm{\hat{k}}$ plane perpendicular to the polar axis is shown in Fig.~\ref{fig:K3}(e). 
As seen, this toy model already reproduces some characteristics of the  DFT computations and their parametrization by the phenomenological Rashba form Eq.~\eqref{eq:Lambda}: strongly anisotropic split for the lower bands $n=1$ and $n=2$ with a relative angular shift of $\pi/4$, and an isotropic split for the upper band $n=3$. 
The agreement is particularly good for the $\bar{S}_2$ mode [Fig.~\ref{fig:K3}(b)] and qualitatively correct for the  $\bar{S}_3$ [Fig.~\ref{fig:K3}(c)].
In the case of the Slater $\bar{S}_1$ mode, the figure is rotated
by $\pi/4$. 
Moreover, the relative coupling strength of the second band with respect to the other two bands is different for modes $\bar S_2$ (smallest coupling) and $\bar{S}_3$ (largest coupling) as seen by comparing the red curve in Figs.~\ref{fig:K3}(b) and Fig.~\ref{fig:K3}(c). These polar eigenvector dependent features cannot be captured by the toy model Eq.~\eqref{eq:Huwt}. Indeed, having only one induced hopping parameter, it can not distinguish among different phonon modes except for an overall mode-dependent factor. Moreover, the relative strength of the interactions is fixed. For example, the maximum of the Rashba coupling of the $n=2$ band is close to that of the other two bands [see Fig.~\ref{fig:K3}(e)]. Certainly more parameters should be considered to describe the full mode dependence, but some conclusions about how the anisotropy arises can already be drawn by studying the present toy model.

Applying perturbation theory on the spin-orbit term $\mathcal{H}_\mathrm{SOC}$ [Eq.~\eqref{eq:Hsoc}] with the hopping term $\mathcal{H}_0$ [Eq.~\eqref{eq:H0}] as a perturbation along a particular momentum direction $\bm{\hat{k}}$, we obtain a $\bm{\hat{k}}$-dependent basis for the $j=3/2$ multiplet, and an isotropic basis for the $j=1/2$ doublet. 
Projecting then the polar term Eq.~\eqref{eq:Huwt} into the perturbative basis we recover a Rashba-like interaction term Eq.~\eqref{eq:Husoc}. 
Because the basis with $\mathcal{H}_0$ as a perturbation is $\bm{\hat{k}}$-dependent for the $j=3/2$ multiplet, so is the resulting Rashba-like interaction, and the induced band split in Fig.~\ref{fig:K3}(e) (blue and red curves). On the contrary, the band split of the $j=1/2$ doublet remains isotropic, like in the conventional Rashba model [Eq.~\eqref{eq:dEiso}]. The $\bm{\hat{k}}$-dependence of the perturbative basis explains why there is a strong (weak) angular correction of the $L=3$ cubic harmonic for the bands stemming from the $j=3/2$ ($j=1/2$) multiplet.
 
From the perturbation analysis, we obtain the following expressions for the Rashba split given by the microscopic model Eq.~\eqref{eq:Htotmodel} at linear order in $k$ along two different momenta directions $\bm{\hat{k}}\parallel[100]$ and $\bm{\hat{k}}\parallel [110]$ for the three electronic bands:
\begin{align}
\label{eq:dE1}
    \frac{\delta E_1(k[100],u)}{2kau}&=\frac{4}{3}\frac{\partial t}{\partial u }; \quad& \frac{\delta E_1(k[110],u)}{2kau}&=\frac{2}{3}\frac{\partial t}{\partial u } \left(1-\frac{\delta-12t_4}{\sqrt{\delta^2+48t_4^2}}\right)\xrightarrow{t_4=0} 0\\
    \label{eq:dE2}
    \frac{\delta E_2(k[100],u)}{2kau}&=0; \quad& \frac{\delta E_2(k[110],u)}{2kau}&=\frac{2}{3}\frac{\partial t}{\partial u } \left(1+\frac{\delta-12t_4}{\sqrt{\delta^2+48t_4^2}}\right)\xrightarrow{t_4=0}\frac{4}{3}\frac{\partial t}{\partial u } \\
    \frac{\delta E_3(k[100],u)}{2kau}&=\frac{4}{3}\frac{\partial t}{\partial u }; \quad& \frac{\delta E_3(k[110],u)}{2kau}&=\frac{4}{3}\frac{\partial t}{\partial u } 
    \label{eq:dE3}
\end{align}
Here we have introduced the hopping parameter $\delta=t_1+2t_3-t_2$ appearing in the eigenstates along $\bm{\hat{k}}\parallel [110]$. It is obvious from Eqs.~\eqref{eq:dE1}-\eqref{eq:dE2} that the two lower bands acquire $\bm{\hat{k}}$-dependent splits with different values along $\bm{\hat{k}}\parallel[100]$ and  $\bm{\hat{k}}\parallel [110]$. The upper $n=3$ band [Eq.\eqref{eq:dE3}], on the contrary, remains isotropic, in a conventional Rashba form. One can check that the perturbative expressions Eqs.~\eqref{eq:dE1}-\eqref{eq:dE3} reproduce the numerical result in Fig.~\ref{fig:K3}(e) in the appropriate directions. 
Note that the band split of $n=1$ and $n=2$ is reduced to simple identical expressions with a relative $\pi/4$ rotation, when taking the even-parity inter-orbital hopping term $t_4$ in Eq.~\eqref{eq:tmunu} to zero.

This comparison between the \emph{ab initio} and the toy model band splitting strongly suggests that different microscopic processes with distinct amplitudes are activated for each of the $\bar S_i$ polar modes, and one needs to go beyond the spin-conserving inter-orbital odd-in-$k$ hopping processes we considered here [Eq.~\eqref{eq:Huwt}]. That is, include additional symmetry allowed terms within the $t_{2g}$ manifold with effects such as orbital polarization, spin-flip hopping channels and virtual processes to the $e_g$ manifold in the minimal polar Hamiltonian Eq.~\eqref{eq:Huwt}, in particular for the (experimentally relevant) Slater mode $\bar{S}_1$. These additional terms should allow one to understand the relevant processes that are needed to reproduce the obtained polar mode dependent \emph{ab initio} results: modify the relative amplitudes of the split in the different electronic bands for  $\bar S_2$ and $\bar{S}_3$, and recover the rotated anisotropy for $\bar S_1$.

With strained materials and heterostructures in mind, we note that the anisotropic linear-in-$k$ Rashba result is fragile around the zone-center towards perturbations that break the degeneracy of the $j=3/2$ manifold at $\Gamma$. We illustrate this effect by introducing  a new term in Eq.~\eqref{eq:Htotmodel} that shifts the energy of the $z$-orbital, and hence breaks the degeneracy of $j=3/2$ [Eqs.~\eqref{eq:c3232}-\eqref{eq:c3212}]: 
\begin{equation}
\label{eq:Htet}
    \mathcal{H}_\mathrm{tet}=\Delta\sum_{\mu s} \delta_{\mu,z}c^\dagger_{\mu s} c_{\mu s}.  
\end{equation}
This term can be used to describe effects such as a tetragonal crystal field in the bulk or differential confinement effects in heterostructures~\cite{Khalsa2013,Zhong2013}. It is also allowed in the presence of a polar distortion along $[001]$ like the one we have considered in this work, but it enters as a quadratic coupling $\Delta \propto \bm{u}^2$. 
Figure~\ref{fig:K3}(f) shows the electronic band split of Eq.~\eqref{eq:Htotmodel} at $\frac{ka}{\pi}=0.01$ with the additional term \eqref{eq:Htet} and symmetry breaking parameter $\Delta=\Delta_\mathrm{SOC}/200=2$ meV, a very small fraction of the SOC gap $\Delta_\mathrm{SOC}$. As seen, the coupling of the lowest band $n=1$ (blue line) has been significantly reduced, and the second band $n=2$ has lost most of its anisotropy, approaching the conventional isotropic Rashba model. The effect is large at small momenta, when the splitting of bands in the absence of SOC is comparable to or smaller than $\Delta$, and fades away otherwise.

This result is in agreement with a recent study in tetragonal STO~\cite{gastiasoro2022theory,gastiasoro2022STO}, a system with $\Delta=\Delta_\mathrm{SOC}/1.6$. The pseudospin band split computed from analogous frozen-phonon computations was indeed found to be fairly isotropic for the lowest two bands in the linear-$k$ regime. Our analysis implies one should expect a similar anisotropic Rashba interaction in the $j=3/2$ multiplet of cubic STO.

Equation~\eqref{eq:Lambda} assumes a conventional Rashba linear-in-$k$ form.
As can be seen in the \emph{ab initio} results of Fig.~\ref{fig:linearfit}(a)-(c), the electronic band split grows linearly with momentum in KTO, but only up to a fraction of the inverse cubic lattice constant, typically $ka\sim 0.6$ along the [100] direction. 
In fact, this deviation from conventional linear-in-$k$ Rashba is very well captured by the toy microscopic model in Eq.~\eqref{eq:Htotmodel}, as can be inferred from Fig.~\ref{fig:linearfit}(d). 

In the case of STO, it was explicitly shown in Ref.~\cite{gastiasoro2022STO}
that this effect can be traced back to a competition between spin-orbit and hopping energies, which leads to a momentum-dependent quenching of the orbital angular momentum. Here, this scenario applies very well for $\bm{k}\parallel[100]$ in KTO. Indeed, the $j=3/2$ multiplet at 
the zone center gets strongly split at finite momentum due to the mass mismatch  [Fig.~\ref{fig:bands}(b)]. As a consequence, the orbital angular momentum gets quenched for some bands as shown in  Fig.~\ref{fig:bands}(c)-(d). Here, for each Kramers doublet, we choose the basis so that it diagonalizes the perturbation corresponding to an infinitesimal polarization $\bm{\hat{n}}_p$ along $z$. 
Once the basis is determined we compute the expectation value of the angular momentum of the two degenerate states. 
One finds that the Rashba coupling [Fig.~\ref{fig:linearfit}(d)] scales with the magnitude of the angular momentum {\em projected in the direction perpendicular to $\bm{n}_p$ and $\bm{k}$} and decreases when this ``perpendicular" component of the angular momentum is quenched [Fig.~\ref{fig:bands}(c)]. This is also valid when comparing different directions. For example, the fact that the perpendicular angular momentum of $n=3$ has the same $k\rightarrow 0$ limit for the two directions in Fig.~\ref{fig:bands}(c) (green line) matches the isotropic form of the Rahsba splitting in Fig.~\ref{fig:K3}(e) and Eq.~\eqref{eq:dE3}.

By the same token, an unquenched angular momentum fully parallel to $\bm{k}$ (as for band $n=2$ with  $\bm{k}\parallel[100] $) does not produce a Rashba splitting [red lines in Figs.~\ref{fig:bands}(c)-(d),\ref{fig:linearfit}(d)]. In general, the deviation from the linear-in-$k$ behavior of the Rashba splitting in the toy model is determined by the point at which the perpendicular component of the  angular momentum deviates from constant. In the case of band $n=1$  where the perpendicular angular momentum is constant  for $\bm{k}\parallel[110]$ (blue lines)  the deviation from linearity is due to lattice effects (i.e. $\sin{(k_i a)}$ in Eq.~\eqref{eq:Huwt}).

\subsection{Electron-polar-phonon Hamiltonian}
\label{subsec:el-ph}

In order to derive the electron-polar-phonon Hamiltonian we quantize the atomic displacements in the polar interaction~\eqref{eq:Lambda} by decomposing them into a set of normal modes $\alpha$, 
\begin{equation}
     \mc{H}_u= \frac{1}{\sqrt{\mathcal{N}}}\sum_{n \bm{k} \bm{q}\alpha }  \psi_n^\dagger(\bm{k}+\frac{\bm{q}}{2}) g^\alpha_{n}(\bm{k},\bm{q}) \psi_{n}(\bm{k}-\frac{\bm{q}}{2})\left(\hat{a}_{\bm{q}\alpha}+\hat{a}^\dagger_{-\bm{q}\alpha} \right)
\end{equation}
where $\mathcal{N}$ is the number of unit cells.
The electron-polar-phonon coupling function is in turn also a $2\times 2$ matrix in pseudospin space,
\begin{equation}
\label{eq:gn}
   \mathfrak{g}^\alpha_{n}(\bm{k},\bm{q}) = ka \sqrt{\frac{\hbar}{2m_{u}\omega_{\bm{q}\alpha}}} \left(\tau^\alpha_{n,1}\bm{\hat{k}}+\tau^\alpha_{n,3}\bm{\mathcal{K}}_3(\bm{\hat{k}})\right)\times\bm{\sigma}\cdot \bm{\hat{n}}_p^{\alpha}(\bm{q})
\end{equation}
and has acquired the cubic structure of the polar interaction $\Lambda_n(\bm{k},\bm{q})$ [Eq.~\eqref{eq:Lambda}] with finite $\tau^\alpha_{n,3}$. Here $m_u$ is the atomic mass constant, and $\omega_{\bm{q}\alpha}$ and $\bm{\hat{n}}^\alpha_p(\bm{q})$ are the frequency and polar axis of the $\alpha$ mode, respectively.

In Ref.~\cite{gastiasoro2022STO} we presented a formalism to obtain the electron-phonon matrix elements in Eq.~\eqref{eq:gn} for a general mode $\alpha$, by decomposing the eigenvector of the $\alpha$ mode into the complete $\bar{S}_i$ basis presented in Section~\ref{sec:Si}. The actual eigenvector of the soft FE mode is however hard to determine accurately due to the anharmonic nature of the system~\cite{vogt1995}.
Both neutron~\cite{Harada1970} and hyper-Raman~\cite{Vogt1988} experiments indicate that the soft FE mode is very close to the Slater $\bar S_1$ mode at room temperature, but whether this trend is kept when lowering $T$ is unknown.
So as a first step, we evaluate the electron-polar-phonon matrix element by assuming the polar mode $\alpha$ to be a pure mode: $\bar{S}_1$, $\bar S_2$ or $\bar S_3$. 
Then the Rashba couplings in Eq.~\eqref{eq:gn} become $\tau^\alpha_{n,j}=\sqrt{\frac{m_u}{\mu_{\bar{S}_i}}}\tau^{\bar{S}_i}_{n,j}$ for $j=1,2,3$, where we have introduced the reduced mass of each $\bar S_i$ mode,
\begin{align}
    \mu^{-1}_{\bar S_1}&=\left(m^\mathrm{Ta}\right)^{-1}+\left(3m^\mathrm{O}\right)^{-1} \\
    \mu^{-1}_{\bar S_2}&=\left(m^\mathrm{K}\right)^{-1}+\left(m^\mathrm{Ta}+3m^\mathrm{O}\right)^{-1} \\
    \mu^{-1}_{\bar S_3}&=\left(m^\mathrm{O}\right)^{-1}+\left(2m^\mathrm{O}\right)^{-1}  
\end{align}
and the electron-polar-phonon coupling function becomes,
\begin{equation}
\label{eq:gnSi}
    \mathfrak{g}^\alpha_{n}(\bm{k},\bm{q}) = ka \sqrt{\frac{\hbar}{2\mu_{\bar{S}_i}\omega_{\bm{q}\alpha}}} \left(\tau^{\bar{S}_i}_{n,1}\bm{\hat{k}}+\tau^{\bar{S}_i}_{n,3}\bm{\mathcal{K}}_3(\bm{\hat{k}})\right)\times\bm{\sigma}\cdot \bm{\hat{n}}_p^\alpha(\bm{q})
\end{equation}
Substituting the experimental zone-center frequency of the soft FE mode at low-$T$ $\omega_{\bm{q}\alpha}=\omega_\mathrm{TO}=2.5$ meV~\cite{vogt1995}, we obtain a characteristic zero-point motion length $l_\mathrm{TO}=\sqrt{\frac{\hbar}{2\mu_{\bar{S}_i}\omega_\mathrm{TO}}}$ equal to 0.15 {\AA} for the $\bar S_1$ mode. 
Substituting in Eq.~\eqref{eq:gnSi} the estimated Rashba couplings $\tau^{\bar{S}_i}_{n,1}$ and $\tau^{\bar{S}_i}_{n,3}$ (listed in Table~\ref{tab:coeffs}) we obtain the following electron-TO-phonon coupling function, 
\begin{equation}
\label{eq:gnTO}
    \mathfrak{g}^\mathrm{TO}_{n}(\bm{k},\bm{q})=ka l_\mathrm{TO}\left(\tau^{\bar S_i}_{n,1}\bm{\hat{k}}+\tau^{\bar S_i}_{n,3}\bm{\mathcal{K}}_3(\bm{\hat{k}})\right)\times\bm{\sigma}\cdot \bm{\hat{n}}_p^\mathrm{TO}(\bm{q}). 
\end{equation}
The matrix-elements for the three $\bar{S_i}$ modes are listed in Table~\ref{tab:coeffs}, and the corresponding plot for the $\bar S_i$ modes polarized along $[001]$ in the perpendicular $k_xk_y$-plane is shown in Figs.~\ref{fig:K3}(a)-(c) [maximum value indicated on the right of the $y$-axis]. 

As seen, the coupling to the soft mode is generally weaker in the pure $\bar S_2$ and $\bar S_3$ modes than in the pure $\bar S_1$ mode, 
i.e. $|g^{\bar S_1}_{n,j}|> |g^{\bar S_3}_{n,j}|> |g^{\bar S_2}_{n,j}|$ for $j=1,3$. 
This implies that starting from a pure $\bar{S}_1$ mode in Eq.~\eqref{eq:gnTO} (as suggested by neutron and hyper-Raman experiments~\cite{Harada1970,Vogt1988}), and then including small to intermediate contributions to the eigenvector of the soft TO mode from $\bar S_2$ and $\bar S_3$ modes should not result in substantial modifications of the estimated electron-TO-phonon matrix element.   
This is unlike tetragonal STO, where the enormous Rashba-like coupling to the $\bar{S}_3$ mode had important consequences~\cite{gastiasoro2022STO}. In cubic KTO this mode appears to be less relevant, simplifying the problem of finding the precise form of the eigenvector of the soft FE mode. 

We can now use the estimated electron-phonon coupling function in Eq.~\eqref{eq:gnTO} to estimate the corresponding SC pairing strength. Solving the full anisotropic SC gap equation is beyond the scope of this article. We obtain instead an approximate estimate of the pairing coupling constant by taking the isotropic limit of the Slater electron-phonon matrix element $\frac{g^\mathrm{TO}}{k_Fa}\approx 100$meV [Fig.~\ref{fig:K3}(a)] for an isotropic $s$-wave SC solution. 
The BCS pairing coupling constant is
$\lambda_\mathrm{BCS}=N_F V_{TO}$ with electronic density of states $N_F$ and effective pairing interaction from the coupling to the TO mode $ V_{TO}=\frac{|g^\mathrm{TO}|^2}{\omega_\mathrm{TO}}\approx (4 \ \mathrm{eV}) (k_Fa)^2$. The estimated pairing strength is indeed very close to analogous pairing estimations for STO from a soft Slater mode~\cite{gastiasoro2022theory}. This is because the smaller bare Rashba coupling $\tau^{\bar S_1}$ to the Slater mode in STO is compensated by a softer mode frequency $\omega_{TO}^\mathrm{STO}\approx0.4\omega_{TO}^\mathrm{KTO}$ (both $\tau^{\bar S_1}$ and $\omega_{TO}$ enter quadratically in $V_{TO}$), giving rise to a similar Rashba pairing interaction estimate in both materials. We emphasize that while we are not aware of any experimental reports of SC in doped bulk KTO to date (doping this system is currently difficult from the synthesis side~\cite{wemple1965,sakai2009thermoelectric}), our simple estimates suggest this is a promising venue to be explored.

\section{Conclusions and Outlook}
\label{sec:conclusions}

We have studied the Rashba-like coupling between the three spin-orbit electronic bands and polar modes in cubic bulk KTO.  
Relevant $q=0$ frozen-phonon \emph{ab initio} computations find a very anisotropic Rashba-like linear-in-$k$ electronic band reconstruction of the lowest $j=3/2$ multiplet, while the reconstruction in the highest $j=1/2$ doublet is fairly isotropic [see Figs.~\ref{fig:linearfit}(a)-(c) and Figs.~\ref{fig:K3}(a)-(c)]. The anisotropy of the $j=3/2$ manifold implies that a minimal interaction describing the linear coupling of the two lowest electronic bands to a polar mode is given by Eq.~\eqref{eq:Lambda}, which includes substantial $L=3$ cubic harmonic corrections $\tau^{\bar S_i}_{n,3}$ for $n=1,2$ [see Table~\ref{tab:coeffs}] beyond the conventional $L=1$ isotropic Rashba model Eq.~\eqref{eq:Lambdaiso}. 

The amplitude and structure of the \emph{ab initio} electronic band splitting shows a pronounced dependence on the type of frozen phonon, i.e. on the polar eigenvector. While the qualitative split in the electronic bands is the same for polar modes $\bar S_2$ and $\bar{S}_3$ [Eqs.~\eqref{eq:S2}-\eqref{eq:S3}], the relative amplitudes of the band split is different in both modes [see Figs.~\ref{fig:K3}(b)-(c)]. Moreover, not only the relative strength of the electronic bands is different for the Slater mode $\bar S_1$ [Eq.~\eqref{eq:S1}], but the anisotropy of the $j=3/2$ multiplet appears also rotated by $\pi/4$ with respect of that in modes $\bar{S}_2$ and $\bar{S}_3$.

In order to understand the origin of the anisotropy in the polar Rashba-like interaction we derived a tight-binding electronic model [Eq.~\eqref{eq:tb-model}] for bulk KTO which excellently captures the low-energy part of the DFT band structure [Fig.~\ref{fig:bands}(b)].  
We construct a toy model which considers in addition to the tight-binding model, symmetry allowed odd-parity spin-conserving inter-orbital hopping processes to describe the interaction of the electrons with a polar mode [Eq.~\eqref{eq:Huwt}]. The anisotropy (isotropy) of the $j=3/2$ ($j=1/2$) manifold Rashba-like interaction is reproduced by this toy model [Fig.~\ref{fig:K3}(e) and Eqs.~\eqref{eq:dE1}-\eqref{eq:dE3}]. The origin of the anisotropy can be traced back to the angular $\bm{\hat{k}}$-dependent manner in which the non-interacting electronic hopping term [Eq.~\eqref{eq:H0}] breaks the degeneracy of the SOC $j=3/2$ manifold at $\bm{k}\rightarrow 0$. It can be also understood by how these symmetry breaking terms affect the orbital angular momentum.

Within the toy-model, the $\bm{\hat{k}}$-dependence in the SOC eigenstates results in a $\bm{\hat{k}}$-dependent anisotropic Rashba coupling to the polar modes for the two lowest bands [Eqs.~\eqref{eq:dE1}-\eqref{eq:dE2}] and the orbital angular momentum [Fig.~\ref{fig:bands}(c)]. The highest $j=1/2$ manifold lacks the extra degeneracy and has isotropic SOC eigenstates and orbital angular momentum and develops a correspondingly isotropic Rashba interaction to polar modes [Eq.\eqref{eq:dE3} and green curve in Fig.~\ref{fig:K3}(e)].
Following this argument one should also expect similar anisotropic Rashba interactions in high-$T$ cubic STO and other systems with highly-degenerate SOC multiplets, such as half-Heusler materials~\cite{savary2017}.

While the toy model Eq.~\eqref{eq:Htotmodel} captures the strong anisotropy (isotropy) of the linear-in-$k$ Rashba interaction of the $j=3/2$ ($j=1/2$) manifold shown in \emph{ab initio} computations, it does not capture several of the polar eigenvector dependent features highlighted above. In particular, the rotated anisotropy of the $j=3/2$ multiplet and the strong renormalization of the relative Rashba couplings of the electronic bands for the Slater $\bar{S}_1$ mode are missed [compare Figs.~\ref{fig:K3}(a) and \ref{fig:K3}(e)]. Since experimentally the eigenvector of the soft FE mode in bulk KTO seems to be very close to  $\bar{S}_1$, a minimal model to describe the coupling to this mode should include additional processes beyond the odd-parity inter-orbital processes we considered in Eq.~\eqref{eq:Huwt}.   

Another important feature reproduced rather well by the toy model Eq.~\eqref{eq:Htotmodel} is the $k$ evolution of the electronic band split found by \emph{ab initio} [Figs.~\ref{fig:linearfit}]. That is, the range of validity for the conventional linear-in-$k$ Rashba interaction Eq.~\eqref{eq:Lambda}. The deviation from linearity in the toy model originates from a competition between SOC and hopping energies, which leads to a pronounced dependence of the orbital angular momentum of the electronic bands on electronic momentum [Fig.~\ref{fig:bands}(c)-(d)]. The situation is particularly simple for momenta along the [100] direction where, at high-momenta, the kinetic terms start to dominate and the spin-orbit assisted Rashba interaction is cut-off and dies out.  

KTO has a SOC interaction which is an order of magnitude larger than in STO~\cite{bruno2019band}. Naively one would expect that this larger interaction should manifest in the electron-polar mode coupling strength. However, similar to STO, due to the orbital
degeneracy of the electronic bands without SOC at $\Gamma$ (Eq.~\ref{eq:H0}), total angular momentum eigenstates can be constructed which diagonalize the SOC interaction independently of its strength. As a consequence the linear-in-$\bm k$ Rashba splittings and corresponding matrix elements are independent of the strength of SOC [c.f. Eqs.~\eqref{eq:dE1}-\eqref{eq:dE3}].
The larger SOC in KTO manifests in the range of $\bm k$ in which a significant Rashba coupling [Eq.~\eqref{eq:Husoc}] is found, before the hopping terms cut-off the interaction (the hopping energies $t_i$ and band mismatches are very similar in STO and KTO). Indeed, for STO along the $[110]$ direction the maximum Rashba matrix element is found for $ka\sim 0.4$~\cite{gastiasoro2022STO}.  For KTO this deviation shifts to much larger momentum values where lattice effects enter into play [c.f. Fig.~\ref{fig:linearfit}(a) and Eq.~\eqref{eq:Huwt}]. This suggests that the Rashba pairing mechanism in KTO may support superconductivity at a larger bulk doping than in STO and is an encouragement for high doping attempts. On the other hand, as doping increases other effects such as screening, the hardening of the TO mode and the softening of the LO mode may become important. More experimental and theoretical work is needed to address these issues. 

We have also derived the electron-polar-phonon coupling function and estimated the corresponding matrix-elements for the three $\bar S_i$ polar modes. The strong Rashba coupling anisotropy in the $j=3/2$ inferred from \emph{ab initio} is of course inherited by the electron-polar-phonon coupling function [Eq.~\eqref{eq:gn} and r.h.s of $y$-axis in Figs.~\ref{fig:K3}(a)-(c)]. 
The Slater-Koster fit~\cite{Khalsa2013} used to estimate the electron-phonon matrix element for the Slater mode in Ref.~\cite{liu2022tunable}, $g^\mathrm{TO}/ka=65$ meV, is within the estimates we have presented here using frozen-phonon \emph{ab initio} instead [see r.h.s of $y$-axis in Fig.~\ref{fig:K3}(a)].

The estimated electron-polar-phonon matrix elements suggest a pairing interaction strength similar to that of STO, making superconductivity in bulk KTO a promising venue to be explored. 
In 2D KTO-based heterostructures superconductivity has been reported, with a strong sensitivity of $T_c$ on the crystallographic orientation of KTO~\cite{liu2022tunable}. It is already interesting that the Rashba-like coupling to the polar modes we presented here in the higher cubic symmetry of the bulk is intrinsically anisotropic. 
In heterostructures there will be additional crystallographic orientation dependent strong crystal field effects affecting both the electrons and their interaction with the polar modes. This calls for \emph{ab initio} computations in relevant slab geometries~\cite{shanavas2014}, including appropriate electrostatic boundary conditions~\cite{brumme2014}, to directly address the crystallographic orientation dependence of the Rashba interaction in heterostructures following the approach we presented here.

\section*{Acknowledgements}
We thank A. Bhattacharya, M. R. Norman and N. Spaldin for useful discussions.
We acknowledge financial support from the Italian MIUR through Projects No. PRIN 2017Z8TS5B, and No. 20207ZXT4Z. 
M.N.G. is supported by the Marie Skłodowska-Curie individual fellowship Grant Agreement SILVERPATH No. 893943. 
We acknowledge the CINECA award under the ISCRA initiative Grants No. HP10CCJFWR and No. HP10CPHFAR, for the availability of high-performance computing resources and support.

\emph{Notes added $-$} While drafting this manuscript we learnt of the recent preprint~\cite{esswein2022first} which explores the electron-phonon coupling to the phonon modes in KTO neglecting spin-orbit coupling, and hence not considering the Rashba interaction vertex we explored here. 

\appendix

\section{Computational details}
\label{app:dft}
{\em Ab initio} calculations have been performed within the Density Functional Theory (DFT) using the projector-augmented wave (PAW) method as implemented in VASP \cite{vasp1,vasp2}. We adopted the generalized gradient approximation revised for solids (PBEsol) \cite{pbesol}. A 12$\times$12$\times$12 Monkhorst-Pack grid of $k$-points has been used for Brillouin-zone integration, with a plane-wave cutoff of 520 eV. The cubic structure with $Pm\bar{3}m$ ($O_h$) symmetry has been fully relaxed  until forces were smaller than 1 meV/\AA, resulting in a lattice parameter $a_0=3.99~$\AA, in excellent agreement with experimental data~\cite{Vousden1951}. Spin-orbit coupling has been included for band-structure computations, both for undistorted cubic structure and for distorted (frozen-phonon) ones where atoms have been displaced according to the $\bar{S}_i$ modes defined in Eqs. (\ref{eq:S1})-(\ref{eq:S3}). We considered several amplitudes $u_i$ for each mode, ranging from 0.5$\times$10$^{-3}$ to 10$\times$10$^{-3}$~\AA, in order to identify the linear regime for the electron-polar-phonon coupling (see~\ref{app:u*}).

\section{Linear coupling regime to polar modes}
\label{app:u*}

\begin{figure}[h]
    \centering 
    \includegraphics[width=0.8\linewidth]{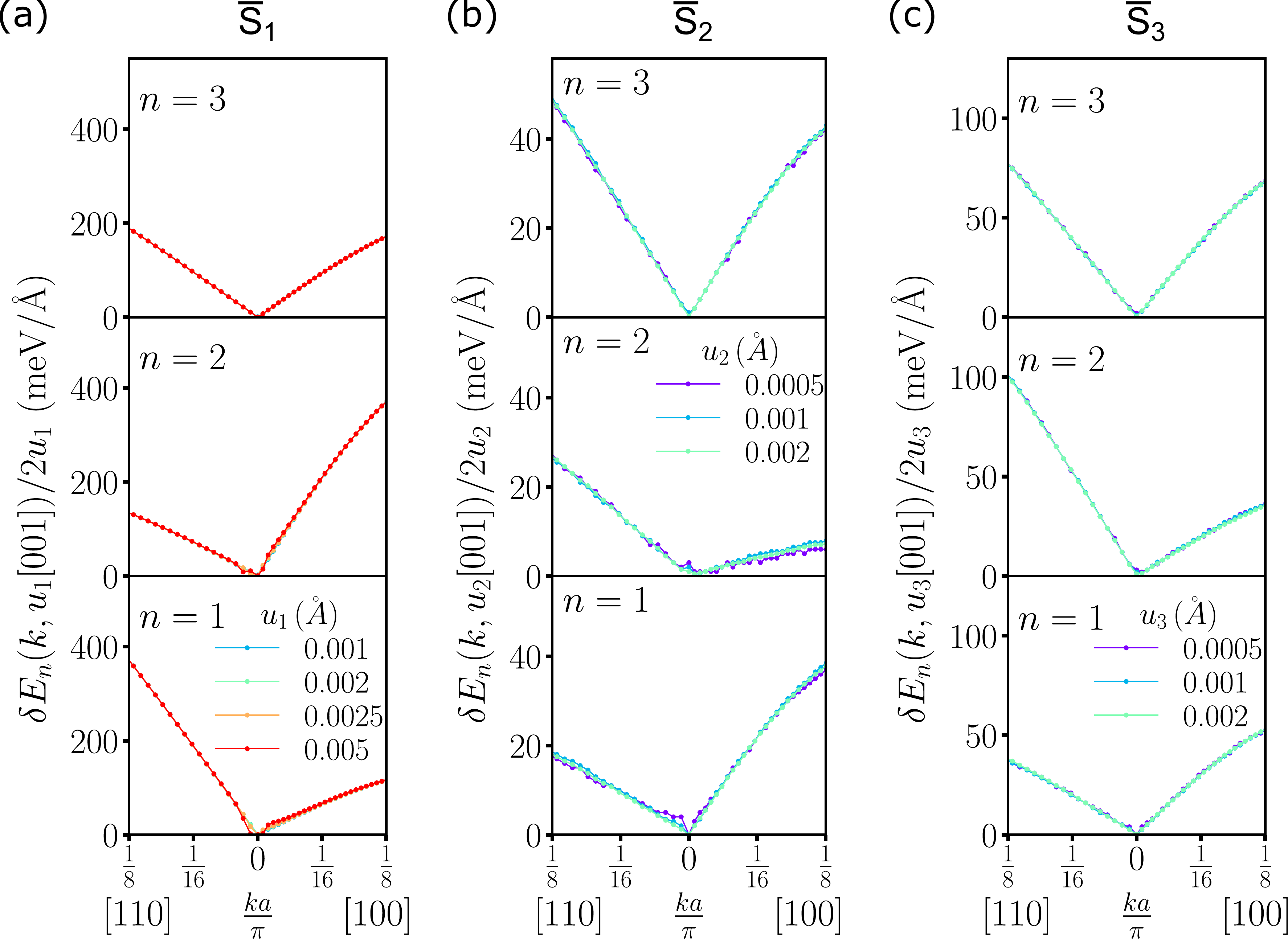}

    \caption{DFT frozen phonon results of the electronic band split $|\delta E_n(\bm{{k}}, \bm{{u}}_i)|$ along the directions $M-\Gamma-X$ normalized by the amplitude $2u_i$ for bands $n=1, 2, 3$ for a polar mode along $[001]$ with eigenvector (a) $\bar S_1$, (b) $\bar S_2$ and (c) $\bar S_3$.
    The critical value $u_i^*$ setting the linear regime for all three bands is $0.002$ {\AA}. 
    The band splitting for modes $\bar{S_2}$ and $\bar{S_3}$ is far smaller than the splitting for $\bar S_1$, causing the observed noise in the data at small amplitudes.}
    \label{fig:linear_ui}
\end{figure}

In this Section we present our DFT results for the three basis polar modes $\bar S_i$ with different $u_i(\bm{q}=\bm{0})$ amplitudes for electronic bands $n=1$, $n=2$ and $n=3$, shown in Fig.~\ref{fig:linear_ui}.
For the $\bar S_1$ mode [Fig.~\ref{fig:linear_ui}(a)] the splitting for all three electronic bands $|\delta E_n(\bm{k})|$ is  of the order of the meV already at $k a/\pi\sim 1/10$ long both $\bm{k}$ directions [110] and [100]. 
A small deviation from the linear regime is found from $u_1=0.002$ {\AA}  to  $u_1=0.0025$ {\AA}.
Therefore, we set the maximum amplitude allowing a linear splitting of the bands for $\bar S_1$ mode to $u_1^*=0.002$ {\AA}.
The same procedure leads to the same regime of linearity for $\bar S_2$ [Fig.~\ref{fig:linear_ui}(b)] and $\bar S_3$ modes [Fig.~\ref{fig:linear_ui}(c)].

Once the linearity near the $\Gamma$ point is guaranteed by the small amplitude of the mode, we perform the linear-in-$k$ fit of the DFT calculations as 
\begin{equation}\label{eq:linear}
    |\delta E_n(\bm{k}, u_i)|/2u_i=ka|\tau^{\bar{S}_i}_{n,DFT}(\bm{\hat{k}})|.
\end{equation} 
The results of the fits are listed in Table~\ref{tab:taudft}. 

\begin{table}[h]
 \caption{Parameter 
 $|\tau^{\bar{S}_i}_{n,DFT}(\bm{\hat{k}})|$ (in meV/\AA) from the linear-in-$k$ fit  
 of Eq.~\eqref{eq:linear} to the frozen-phonon DFT results in Fig.~\ref{fig:linearfit} for polar modes $\bar S_1,\,\bar S_2,\,\bar S_3$. In all cases the polar axis $\bm{\hat{n}}_p\parallel [001]$ and $\bm{\hat{k}}\perp \bm{\hat{n}}_p$.}
     \begin{tabular}{c|ccc|ccc|ccc}
         & & $\bar S_1$ & & & $\bar S_2$ & & & $\bar S_3$ \\
         $\bm{\hat{k}}$  & $n=1$ &$n=2$ &$n=3$ & $n=1$ &$n=2$ &$n=3$ &  $n=1$ &$n=2$ &$n=3$  \\
         \hline
         $[100]$  &  316 & 1042 & 472 & 109 & 20 & 120 & 151 & 95 & 192 \\
         $[110]$  &  940 & 360 & 478 & 47 & 69 & 125 & 99 & 260 & 197
    \end{tabular}
    \label{tab:taudft}
\end{table}

\bibliography{biblio}

\end{document}